\documentclass[11pt,a4paper]{article}

\usepackage[utf8]{inputenc}
\usepackage[T1]{fontenc}
\usepackage[english]{babel}

\usepackage{fullpage}
\usepackage{setspace}
\usepackage{geometry}
\usepackage{amsmath, amssymb, amsfonts, amsthm}
\usepackage{amsopn}
\theoremstyle{plain}

\usepackage{graphicx}
\usepackage{subcaption}

\usepackage{booktabs, tabularx}
\usepackage{multirow}
\usepackage{array, longtable}
\usepackage{rotating}
\usepackage[flushleft]{threeparttable}
\newcolumntype{L}[1]{>{\raggedright\let\newline\\\arraybackslash\hspace{0pt}}m{#1}}
\newcolumntype{C}[1]{>{\centering\let\newline\\\arraybackslash\hspace{0pt}}m{#1}}
\newcolumntype{R}[1]{>{\raggedleft\let\newline\\\arraybackslash\hspace{0pt}}m{#1}}
\usepackage[table]{xcolor}

\usepackage{enumitem}
\usepackage{comment}
\usepackage[toc,page]{appendix}
\usepackage[bottom]{footmisc}

\usepackage{natbib}

\usepackage[colorlinks=true, linkcolor=blue, urlcolor=blue,
            citecolor=blue, anchorcolor=blue]{hyperref}

\newcommand{\GPR}{\textit{GPR}}
\newcommand{\CA}{\textsc{CausalAlpha}}
\newcommand{\indep}{\perp\!\!\!\perp}
\newcommand{\nindep}{\not\!\perp\!\!\!\perp}

\title{CausalAlpha: A Real-Time Geopolitical Risk Index from OSINT
       Channels for Causal Discovery in Financial Markets}
\author{Andr\'{e}s Azqueta-Gavald\'{o}n\thanks{Independent Researcher.
        \texttt{andres.azqueta@gmail.com}}
        \and
        Borja Ureta\thanks{Independent Researcher. \texttt{borjaureta@gmail.com}}}
\date{ \today}

\begin{document}
\maketitle

\begin{abstract}
We introduce \CA{}, an open-source framework that constructs a
high-frequency Geopolitical Risk (\GPR{}) index from Telegram OSINT
channels using natural language processing, and applies causal
discovery methods to identify the directed causal structure between
geopolitical uncertainty and financial market variables. Unlike
standard sentiment indices or Granger-causality approaches, \CA{}
employs the Peter--Clark (PC) algorithm to recover the directed
acyclic graph (DAG) of causal dependencies between five
category-specific \GPR{} indicators and a set of financial variables
spanning commodity prices, equity indices, and credit instruments,
estimated across four DAG specifications and three significance
levels with 500 block-bootstrap resamples. Two findings emerge as
globally robust across all DAG specifications at $\alpha = 0.10$:
political instability and energy media coverage independently and
causally precede conflict coverage, establishing conflict as the
primary causal sink of geopolitical narrative escalation in
real-time OSINT channels. At the strictest significance level
($\alpha = 0.05$), conflict coverage causally precedes energy sector
equity returns ($\Delta$XLE), consistent with geopolitical
escalation transmitting to energy markets. A Structural VAR on the
core macro panel confirms that dynamic transmission from
geopolitical NLP signals to financial market prices is statistically
weak at daily frequency, suggesting that geopolitical news signals
operate primarily within the media narrative system. The framework
is deployed as a production application on Google Cloud Run with
automated data collection and index construction, representing a
step toward real-time macrofinancial risk monitoring using OSINT.

\end{abstract}

\clearpage

% ============================================================
\section{Introduction}
\label{sec:intro}
% ============================================================

The measurement of geopolitical risk (\GPR{}) and its transmission to
financial markets has attracted growing attention following the seminal
contribution of \citet{CaldaraIacoviello2022}. \GPR{} captures the risk
associated with wars, terrorist acts, and tensions between states that
affect the normal and peaceful course of international relations, and
has traditionally been measured using newspaper-based sources.
\citet{CaldaraIacoviello2022} document that stock returns experience a
short-lived but significant drop in response to geopolitical
escalation---an effect that varies substantially across industries,
with the defense sector experiencing positive excess returns while
sectors exposed to the broader economy, such as steelworks and mining,
bear negative ones.

However, traditional media-based indices face a fundamental limitation:
newspapers are slow, curated, and filtered through editorial processes
that introduce a systematic lag between geopolitical events and their
measurement. The proliferation of real-time social media channels---and
in particular encrypted messaging platforms such as Telegram---has
created a new layer of open-source intelligence (OSINT) that reflects
geopolitical developments with minimal delay \citep{SoulaEtAl2024}. A
striking illustration is provided by the Russia--Ukraine conflict, where
intelligence on Russian troops' border movements and military plans
circulated on social media in the hours before President Putin announced
the full-scale invasion \citep{Karalis2024}.

This paper makes three contributions. First, we construct a novel
high-frequency \GPR{} index from Telegram OSINT channels using a
keyword-based NLP pipeline updated daily, covering five geopolitical
risk dimensions: conflict, political instability, energy security,
financial stress, and trade disruption. Second, we apply causal
discovery---specifically the PC algorithm of
\citet{SpirtesEtAl2000}---to identify the directed causal structure
between our \GPR{} indicators and a broad set of financial market
variables spanning commodities, FX rates, equity sector ETFs, and
credit instruments. Unlike Granger causality, the PC algorithm
conditions independence tests on the full variable set, avoiding
spurious inference due to omitted common causes. Third, we release an
open-source production system (\CA{}) deployed on Google Cloud Run
that automates data collection, index construction, causal graph
estimation across multiple DAG specifications and significance levels,
and weekly reporting.

Our results reveal a consistent causal architecture within the
geopolitical media narrative system: political instability and energy
media coverage causally precede conflict coverage across all four DAG
specifications, while conflict coverage causally precedes energy sector
equity returns at the strictest significance threshold ($\alpha = 0.05$).
Dynamic transmission from geopolitical NLP signals to financial market
prices is statistically weak at daily frequency, consistent with
geopolitical information being incorporated into markets faster than it
reaches the media cycle. While \citet{VerduzcoZanetti2026} exploit
high-frequency oil futures responses to spikes in the newspaper-based
GeoThreats index of \citet{CaldaraIacoviello2022} to identify
geopolitical oil price shocks, we construct our indicators directly
from real-time Telegram OSINT channels and apply causal discovery
rather than a pre-specified structural VAR, recovering the full directed
graph of dependencies without imposing prior ordering restrictions.

The paper proceeds as follows. Section~\ref{sec:literature} reviews
the related literature. Section~\ref{sec:data} describes the data and
\GPR{} index construction. Section~\ref{sec:methodology} presents the
methodology. Section~\ref{sec:results} reports empirical results.
Section~\ref{sec:robustness} discusses robustness.
Section~\ref{sec:conclusion} concludes.

% ============================================================
\section{Related Literature}
\label{sec:literature}
% ============================================================

\subsection{Geopolitical Risk Measurement}

The seminal contribution of \citet{CaldaraIacoviello2022} formalised
the measurement of geopolitical risk through automated keyword-based
searches of newspaper archives. Their monthly \GPR{} index has been
widely applied to study the effects of geopolitical uncertainty on
macroeconomic outcomes, investment, and financial markets. The
Economic Policy Uncertainty (EPU) index of \citet{BakerBloomDavis2016}
follows a similar approach, measuring uncertainty about which economic
policies will be implemented and when. Several subsequent contributions
have extended these methodologies using topic modelling
\citep{azqueta2023sources, azqueta2017developing} and large language
models \citep{ghomiraui}, demonstrating that richer text representations
can sharpen the measurement of uncertainty.

\subsection{Social Media as Financial Signal}

\citet{BollenEtAl2011} demonstrated that Twitter mood predicted DJIA
movements up to four days in advance, establishing social media as a
source of financially relevant information. More recently, large
language models have been applied to extract sentiment and
forward-looking signals from social media at scale. However, the use
of Telegram as a geopolitical OSINT source---where real-time reporting
by conflict monitors, military analysts, and investigative journalists
frequently precedes traditional media coverage---remains understudied
in the academic literature.

\subsection{Causal Discovery in Economics and Finance}

The causal discovery literature, following \citet{Pearl2009} and
\citet{SpirtesEtAl2000}, provides a principled framework for
recovering causal structure from observational data without requiring
experimental variation. Applications in economics include
\citet{MonetaEtAl2013} on monetary policy transmission and
\citet{HyvarinenEtAl2010} on structural VAR identification using
non-Gaussianity. A key advantage of constraint-based methods such as
the PC algorithm over pairwise Granger causality is that independence
tests are conditioned on the full variable set, ruling out spurious
causal inference due to common causes. This property is particularly
valuable in our setting, where financial variables share common global
risk drivers.

% ============================================================
\section{Data and GPR Index Construction}
\label{sec:data}
% ============================================================

\subsection{Telegram Data Collection}

Our Telegram corpus is constructed from six OSINT and independent
news channels selected for their coverage of geopolitical events,
conflict monitoring, and international security. Channels were chosen
to balance neutral open-source intelligence sources with editorially
independent journalism, deliberately excluding sources identified as
pro-Kremlin or disinformation vectors.\footnote{IntelSlava Z, a
channel widely cited in earlier Telegram-based studies, was excluded
on grounds of documented pro-Russian bias and disinformation
\citep{SoulaEtAl2024}.} The data collection pipeline uses the Telethon
Python library to scrape channel messages in real time, stored in
Google Firestore. Table~\ref{tab:channels} describes the corpus.

\begin{table}[!htbp]
\centering
\caption{Telegram channels in the \CA{} corpus}
\label{tab:channels}
\begin{tabular}{L{3cm} L{3.5cm} L{4cm} L{3cm}}
\toprule
\textbf{Channel} & \textbf{Handle} & \textbf{Focus} &
\textbf{Bias profile} \\
\midrule
Bellingcat       & @bellingcat                  & OSINT investigations       & Neutral \\
WarTranslated    & @wartranslated               & Frontline Ukraine          & Neutral \\
Reuters World    & @ReutersWorldChannel         & International wire service & Neutral \\
BBC World        & @BBCWorld                    & International news         & Neutral \\
Kyiv Independent & @KyivIndependent\_official   & Ukraine journalism         & Pro-Ukraine \\
Meduza           & @meduzaio                    & Russian independent media  & Anti-Kremlin \\
\bottomrule
\multicolumn{4}{l}{\footnotesize\textit{Note:} All channels post primarily in English.
Total messages processed: 19,543 over the sample period.}
\end{tabular}
\end{table}

Note that the corpus is heavily oriented toward the Russia--Ukraine
conflict and associated geopolitical developments. Hence results
should be interpreted as pertaining to a \textit{conflict-intensive
OSINT sample} rather than geopolitical risk in general. Channels
with broader economic and financial coverage would be needed
to generalise findings to non-conflict geopolitical risk
dimensions such as trade policy uncertainty or sovereign
debt stress.

\subsection{GPR Index Construction}

Rather than constructing a single aggregate \GPR{} index, \CA{}
produces five category-specific indicators that capture distinct
dimensions of geopolitical risk. For each category $c$ and day $t$,
the raw indicator is defined as:
\begin{equation}
\label{eq:gpr}
\widehat{\mathit{GPR}}_{c,t}
  = \frac{\text{Messages on day }t\text{ containing category-}c\text{ keywords}}
         {\text{Total messages on day }t}
  \times 100.
\end{equation}
To smooth daily noise while preserving weekly variation, we compute a
7-day rolling share $\overline{\mathit{GPR}}_{c,t} =
\frac{1}{7}\sum_{k=0}^{6}\widehat{\mathit{GPR}}_{c,t-k}$, which
constitutes the indicator entering the causal analysis.

The five categories and their average message shares over the sample
period are reported in Table~\ref{tab:indicators}. The keyword
dictionary is provided in Appendix~\ref{app:keywords}.

\begin{table}[!htbp]
\centering
\caption{CausalAlpha NLP indicators}
\label{tab:indicators}
\begin{tabular}{llr}
\toprule
\textbf{Indicator} & \textbf{Description} & \textbf{Avg.\ message share} \\
\midrule
Conflict  & Military conflict, terrorism, cyberwarfare          & 64.3\% \\
Political & Political instability, elections, civil unrest      & 17.6\% \\
Energy    & Energy security, pipeline disruption, supply risk   & 11.0\% \\
Trade     & Trade disruption, supply chains, export controls    &  4.5\% \\
Financial & Financial stress, sovereign debt, currency crises   &  0.2\% \\
\bottomrule
\end{tabular}
\begin{tablenotes}
\item \textit{Note:} Average daily message share computed over 19,543 messages across six Telegram channels, April 2025--April 2026. Shares sum to more than 100\% as a single message can activate multiple categories.
\end{tablenotes}
\end{table}

\begin{figure}[!htbp]
\centering
\includegraphics[width=\linewidth,height=0.55\textheight,
    keepaspectratio]{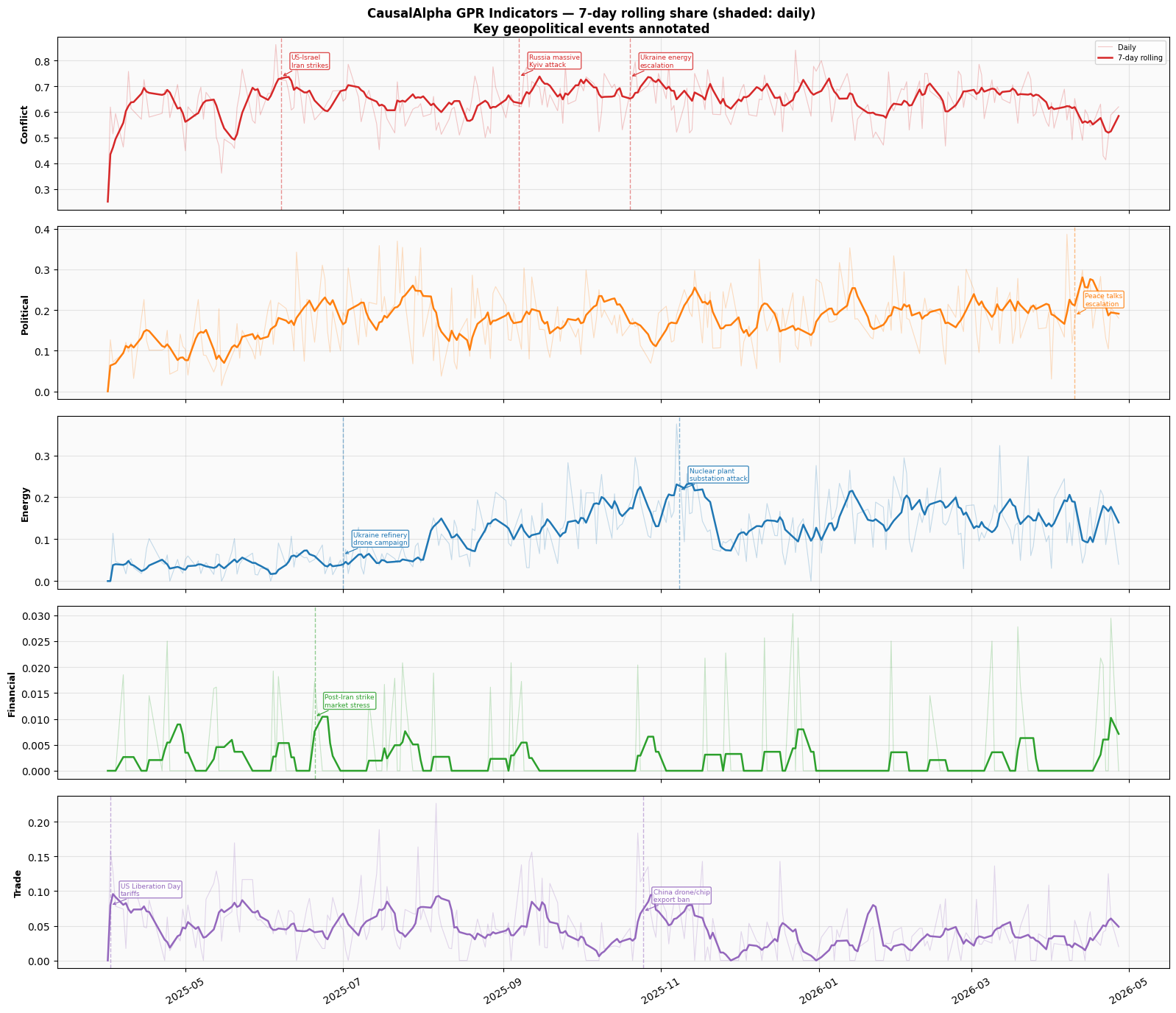}
\caption{CausalAlpha \GPR{} indicators, daily and 7-day rolling 
share, April 2025--April 2026. Shaded lines show daily raw keyword 
shares; bold lines show 7-day rolling means (equation~\ref{eq:gpr}). 
Dashed vertical lines indicate key geopolitical events: 
US--Israel strikes on Iranian nuclear facilities (June 2025), 
Russia's massive attack on Kyiv (September 2025), 
intensification of Ukrainian drone strikes on Russian energy 
infrastructure (July--November 2025), 
the US Liberation Day tariff announcements (April 2025), 
and China's export ban on drone components and semiconductor chips 
(October 2025).}
\label{fig:gpr_ts}
\end{figure}

Figure~\ref{fig:gpr_ts} plots the five \CA{} \GPR{} indicators over 
the sample period. Several features of the series are worth noting. 
The Conflict indicator is the dominant signal throughout, reflecting 
the channel composition: Bellingcat, WarTranslated, and the Kyiv 
Independent focus heavily on the Russia--Ukraine conflict, which 
generates sustained conflict-related coverage at 55--70\% of daily 
messages. The Trade indicator exhibits two sharp spikes --- in April 
2025 following the US Liberation Day tariff announcements, and in 
October 2025 coinciding with China's sequential export bans on drone 
components and semiconductor chips --- consistent with these events 
generating discrete, short-lived surges in trade disruption coverage.

The most economically meaningful feature of the series is the 
structural break in the Energy indicator. Prior to July 2025 the 
indicator fluctuates at 2--5\% of daily messages. From July 2025 
onwards it permanently shifts to 10--20\%, coinciding with the 
intensification of Ukrainian drone strikes that forced approximately 
40\% of Russia's oil refining capacity offline by October 2025 
\citep{RussiaMatters2025}. The November 2025 peak in the Energy 
indicator reflects Russia's 8 November attack on substations 
supplying the Khmelnytskyi and Rivne nuclear power plants 
\citep{ACLED2025}. This step-change is not a data artefact --- it 
reflects a genuine structural shift in the geopolitical news cycle 
in which energy infrastructure became the primary contested domain 
of the conflict. The Financial indicator remains sparse throughout, 
consistent with the channel composition being oriented toward 
conflict monitoring rather than financial reporting.

\subsection{Financial Market Variables}

We collect daily closing prices for 16 financial market variables from
Yahoo Finance, spanning four asset classes: commodities (Brent crude
oil, gold, wheat, copper, natural gas, silver), FX rates (USD/RUB,
USD/TRY, USD/SAR, DXY), equity indices and sector ETFs (S\&P 500,
energy sector XLE, defense sector ITA, financials sector XLF), and
credit instruments (EM sovereign bonds EMB, high-yield bonds HYG). The
CBOE Volatility Index (VIX) serves as a direct measure of financial
market uncertainty. All price-level series are first-differenced before
estimation to ensure stationarity; VIX enters in levels. 

% ============================================================
\section{Methodology}
\label{sec:methodology}
% ============================================================

\subsection{The PC Algorithm for Causal Discovery}

The PC algorithm \citep{SpirtesEtAl2000} is a constraint-based causal
discovery method that recovers the Markov equivalence class of the
true causal DAG from observational data. We apply it to a variable set
$\mathbf{V}$ comprising the five NLP indicators and a domain-specific
subset of financial market variables (detailed below). The algorithm
proceeds in two stages.

\paragraph{Stage 1---Skeleton estimation.}
Starting from a complete undirected graph over $\mathbf{V}$, edges are
removed between variable pairs $(X_i, X_j)$ if there exists a
conditioning set $\mathbf{S} \subseteq \mathbf{V} \setminus
\{X_i, X_j\}$ such that $X_i \indep X_j \mid \mathbf{S}$.
Conditional independence is tested using Fisher's Z-test, appropriate
for approximately Gaussian data after first-differencing.

We note that the Financial indicator (0.2\% average message share)
and, to a lesser extent, the Trade indicator (4.5\%) are sparse and
right-skewed, which may violate the approximate Gaussianity assumption
underlying Fisher's Z-test. Any edge involving the Financial variable
should therefore be interpreted with additional caution. A rank-based
alternative such as the kernel conditional independence test
\citep{ZhangEtAl2012} would be more robust to non-Gaussianity but is
computationally prohibitive at our sample size; we flag this as a
direction for future work.

\paragraph{Stage 2---Orientation.}
V-structures (colliders of the form $X_i \to X_k \leftarrow X_j$ with
$X_i \nindep X_j \mid X_k$) are identified and remaining edges are
oriented using Meek's propagation rules \citep{Meek1995}, yielding a
Completed Partially Directed Acyclic Graph (CPDAG).

\subsection{Multi-DAG Specification and Alpha Sweep}

To assess the robustness of causal structure across market domains and
significance thresholds, we estimate the PC algorithm across a
$4 \times 3$ sweep: four DAG specifications times three significance
levels $\alpha \in \{0.05, 0.10, 0.15\}$ to report the highest level of significance, yielding 12 PC estimations per analysis run. Each DAG specification conditions on the same five NLP indicators but a different domain-specific market subset:

\begin{itemize}[itemsep=2pt]
  \item \textbf{DAG 1 (Core macro):} VIX, $\Delta$Brent, $\Delta$Gold,
        $\Delta$SP500: \textbf{9 nodes}.
  \item \textbf{DAG 2 (Commodities):} VIX, $\Delta$Brent, $\Delta$Wheat, $\Delta$Copper, $\Delta$Natural Gas: \textbf{10 nodes}.
  \item \textbf{DAG 3 (Credit \& FX):} VIX, $\Delta$EM Bonds, $\Delta$HY Bonds, 
    $\Delta$USD/RUB, $\Delta$USD/TRY: \textbf{10 nodes}.
  \item \textbf{DAG 4 (Sector ETFs):} VIX, $\Delta$SP500, $\Delta$XLE,
        $\Delta$ITA, $\Delta$XLF: \textbf{10 nodes}.
\end{itemize}

\subsection{Block-Bootstrap Confidence}

To quantify the sampling uncertainty of each identified edge, we
implement a block bootstrap with 500 resamples per (DAG, $\alpha$)
cell. Each resample is constructed from $\lceil N / 7 \rceil$ randomly
drawn contiguous blocks of seven observations, preserving the
autocorrelation structure induced by the 7-day rolling window of the
NLP indicators. The bootstrap probability of an edge is defined as the
fraction of resamples in which that directed edge appears in the PC
output. Failed resamples contribute a conservative ``edge absent''
vote, ensuring bootstrap probabilities are not inflated by numerical
failures.

\subsection{Structural VAR Validation}

To obtain dynamic impulse responses, we estimate a Structural VAR
(SVAR) of order $p$ on the DAG 1 (core macro) sub-panel:
\begin{equation}
\mathbf{y}_t = \mathbf{c} + \sum_{k=1}^{p} \mathbf{A}_k
\mathbf{y}_{t-k} + \mathbf{u}_t, \qquad
\mathbf{u}_t \sim \mathcal{N}(\mathbf{0}, \boldsymbol{\Sigma}),
\end{equation}
where $\mathbf{y}_t$ stacks the DAG 1 variables. The structural form
is identified via a Cholesky decomposition using the recursive ordering
derived from the PC-algorithm CPDAG: when a topological sort of the
directed edges is available, it is used directly; otherwise we apply
the fallback ordering VIX $\to$ Conflict $\to$ Brent $\to$ Gold
specified in the architecture. We report cumulative impulse response
functions (IRFs) with 95\% Monte Carlo confidence intervals.

% ============================================================
\section{Results}
\label{sec:results}
% ============================================================

\subsection{Cross-DAG Robust Edges}

Table~\ref{tab:robust_edges} reports the directed edges satisfying
our robustness criterion: present in all eligible DAG specifications
at $\alpha = 0.10$. We use $\alpha = 0.05$ as a supplementary
sensitivity check; results at $\alpha = 0.15$ are reported in
Appendix~\ref{app:alpha_sweep}. Panel A reports three globally
robust edges and Panel B reports three domain-specific edges
present in exactly one eligible DAG specification.

\begin{table}[!htbp]
\centering
\caption{Robust causal edges}
\label{tab:robust_edges}
\begin{threeparttable}
\begin{tabular}{llccc}
\toprule
\textbf{Source} & \textbf{Target} & \textbf{DAGs present} &
\textbf{Bootstrap prob.} & \textbf{$\alpha$=0.05} \\
\midrule
\multicolumn{5}{l}{\textit{Panel A: Globally robust (all eligible DAGs at $\alpha$=0.10)}} \\[2pt]
Political           & Conflict  & 4/4 & 0.330 & $\times$ \\
Energy              & Conflict  & 4/4 & 0.252 & $\times$ \\
$\Delta$ SP500      & VIX       & 2/2 & 0.336 & $\times$ \\
\midrule
\multicolumn{5}{l}{\textit{Panel B: Domain-specific (1 eligible DAG at $\alpha$=0.10)}} \\[2pt]
$\Delta$ XLE Energy & Conflict  & 1/1 (DAG 4) & 0.528 & $\times$ \\
$\Delta$ Gold       & Financial & 1/1 (DAG 1) & 0.410 & $\times$ \\
$\Delta$ EM Bonds   & VIX       & 1/1 (DAG 3) & 0.330 & \checkmark \\
\bottomrule
\end{tabular}
\begin{tablenotes}
\item \textit{Note:} Bootstrap probability = fraction of 500
block-bootstrap resamples (block size = 7 days) in which the directed
edge appeared. Failed resamples count as edge absent (conservative
denominator). The $\alpha$=0.05 column indicates whether the edge
survives at the stricter significance level
(\checkmark\ = survives; $\times$ = drops out).
For Panel A edges, bootstrap probabilities are averaged across all
eligible DAG specifications. $\Delta$ denotes first-differenced
price series.
\end{tablenotes}
\end{threeparttable}
\end{table}

Table~\ref{tab:robust_edges} reports bootstrap probabilities and
sensitivity to significance level for all edges identified at
$\alpha = 0.10$. Our robustness criterion requires presence in
all eligible DAG specifications at $\alpha = 0.10$ and
$\alpha = 0.15$; we use $\alpha = 0.05$ as a supplementary
sensitivity check rather than a component of the criterion.
Bootstrap probabilities range from 0.252 to 0.528, reflecting
the moderate sample size of $N = 267$ observations (approximately
38 effective bootstrap blocks of size 7). The two globally robust
edges in Panel A --- Political $\to$ Conflict and Energy $\to$
Conflict --- drop out at $\alpha = 0.05$, indicating that they
reflect moderate rather than strong causal evidence; the
cross-DAG criterion is the primary robustness device, with
bootstrap probabilities serving as a secondary diagnostic.

One domain-specific edges in Panel B do survive at $\alpha = 0.05$:
$\Delta$EM Bonds $\to$ VIX and a bootstrap probability of 0.330. This indicates that stress in the emerging market bonds ultimately originates imbalances in global financial markets. The edge between Conflict and $\Delta$XLE Energy carries the highest bootstrap probability 0.528 but does not survive the 0.05 significance interval. Recall that Energy Select Sector SPDR Fund (XLE) is the largest and most heavily traded exchange-traded fund focusing on the U.S. energy sector. Hence, changes in this ETF anticipate escalation of conflicts. This can be the case given that markets tend to anticipate large scale conflicts (see \citet{chadefaux2017market}). 

Looking at the four DAGs, two findings stand out. First, \textbf{Conflict media coverage} emerges as the primary causal sink in the geopolitical risk system, receiving causal inputs from Political and Energy in all four DAG specifications. The convergence of these three causal channels onto a single target suggests that conflict coverage in OSINT channels is not a primitive signal but rather an aggregation of geopolitical and financial 
pressures that have already manifested in other dimensions of the news 
cycle. A near-robust pattern --- present in three of four DAG specifications
but absent in DAG 4 --- is VIX $\to$ Conflict, suggesting that 
financial market stress \textit{may} amplify media attention to
conflict. This direction should be treated with caution: it is not
globally robust, fails at $\alpha = 0.05$, and 
instead suggests a feedback mechanism in which market stress amplifies 
media attention to conflict. This is consistent with 
\citet{CaldaraIacoviello2022}, who document that geopolitical risk is 
partly endogenous to financial market conditions, and with 
\citet{BakerBloomDavis2016}, who show that uncertainty indices respond 
to financial market volatility as well as generating it. 

The finding that \textbf{Political $\to$ Conflict} and \textbf{Energy $\to$ Conflict} 
are also globally robust is consistent with the political economy 
literature: \citet{AlesinaPerotti1996} document that political instability 
is systematically preceded by deteriorating economic conditions and cuasing violent phenomena, and \citet{ChowdhuryEtAl2025} show that energy supply disruptions generate 
conflict escalation in energy-dependent regions. 

Second, the most novel result of the analysis concerns the relationship 
between Conflict media coverage and $\Delta$XLE Energy identified in 
DAG 4 (Panel B, Table~\ref{tab:robust_edges}), which carries the highest 
bootstrap probability of all reported edges at 0.528. At $\alpha = 0.05$ 
--- the preferred orientation on statistical and economic grounds --- 
Conflict causally precedes $\Delta$XLE Energy, consistent with 
geopolitical escalation transmitting to energy sector equity returns. The bootstrap stability of this edge (0.528) indicates that the Conflict--$\Delta$XLE Energy relationship is the most resampling-stable finding in the analysis. The preferred $\alpha = 0.05$ orientation --- Conflict $\to$ $\Delta$XLE Energy --- is consistent with geopolitical escalation transmitting to energy equity markets before it fully propagates through the broader media cycle. This result is consistent with the broader  price discovery literature documenting that forward-looking financial instruments incorporate information before it diffuses to the broader 
public \citep{Fama1970}, and with \citet{VerduzcoZanetti2026}, who show 
using a high-frequency Proxy VAR that geopolitical oil price shocks 
transmit most strongly to energy-intensive commodities, with European 
natural gas experiencing the largest price response among all commodity 
markets. Our results complement their finding by suggesting that the 
reverse channel also operates: energy equity markets are not passive 
recipients of geopolitical shocks but active price discovery mechanisms 
that lead the media narrative. This is consistent with informed trading 
in energy futures prior to conflict events, and with the episode 
documented by \citet{PolymarketIran2026}, in which six wallets earned approximately \$1.2 million by correctly betting on prediction market contracts tied to Iranian military
developments before the strikes appeared in news coverage. Crucially, 
\citet{VerduzcoZanetti2026} identify their geopolitical shocks using 
the newspaper-based GeoThreats index of \citet{CaldaraIacoviello2022}; 
our OSINT-based indicators, which capture geopolitical signals with 
less editorial lag, nonetheless appear downstream of energy equity 
markets, suggesting that the chosen real-time Telegram channels do not fully 
close the information gap between financial markets and public media.

\subsection{DAG-Specific Results}
\label{sec:dag_results} 

\begin{figure}[!htbp]
\centering
\includegraphics[width=\linewidth,height=0.45\textheight,
    keepaspectratio]{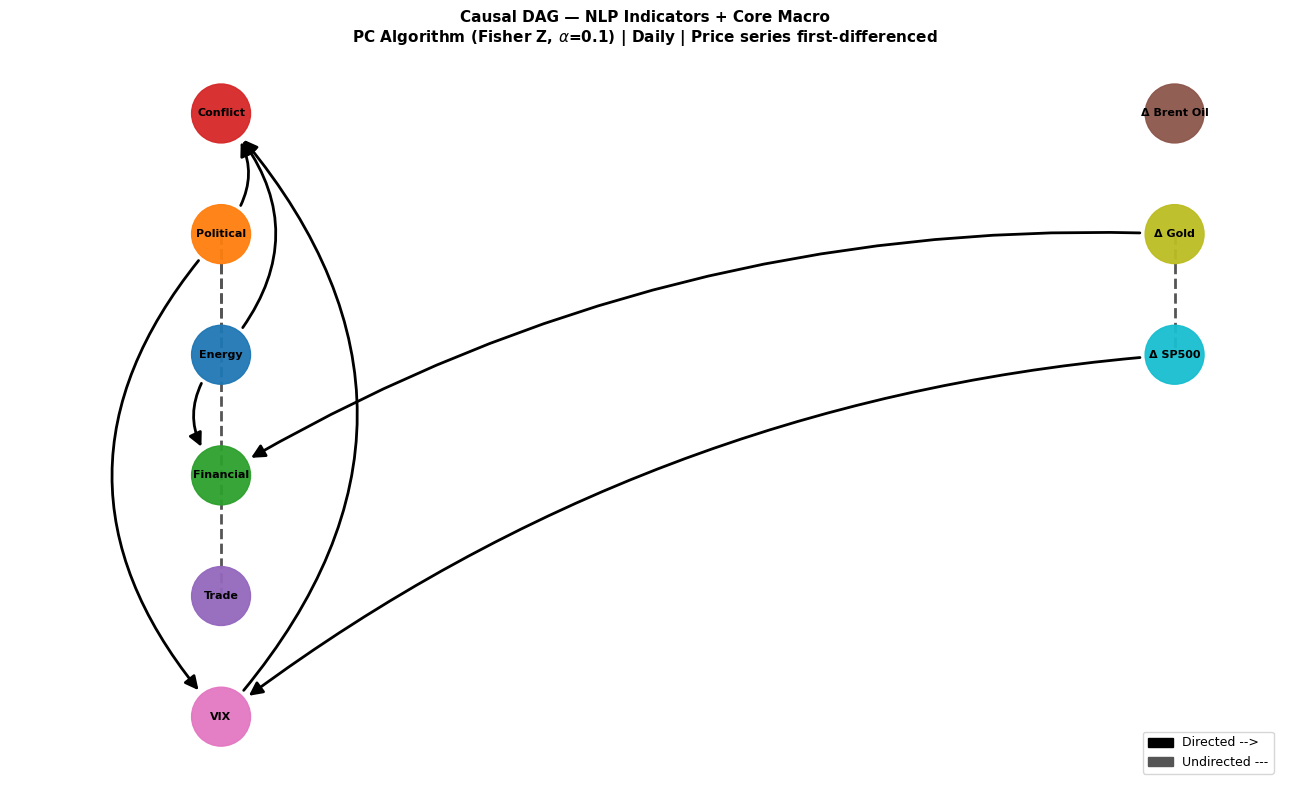}
\caption{Causal DAG 1 --- NLP indicators + core macro variables
(VIX, $\Delta$Brent, $\Delta$Gold, $\Delta$SP500) and the VIX.
PC algorithm, Fisher Z, $\alpha = 0.10$.
Solid arrows: directed edges. Dashed lines: undirected edges.
$N = 267$ daily observations.}
\label{fig:dag_core}
\end{figure}

\begin{figure}[!htbp]
\centering
\includegraphics[width=\linewidth,height=0.45\textheight,
    keepaspectratio]{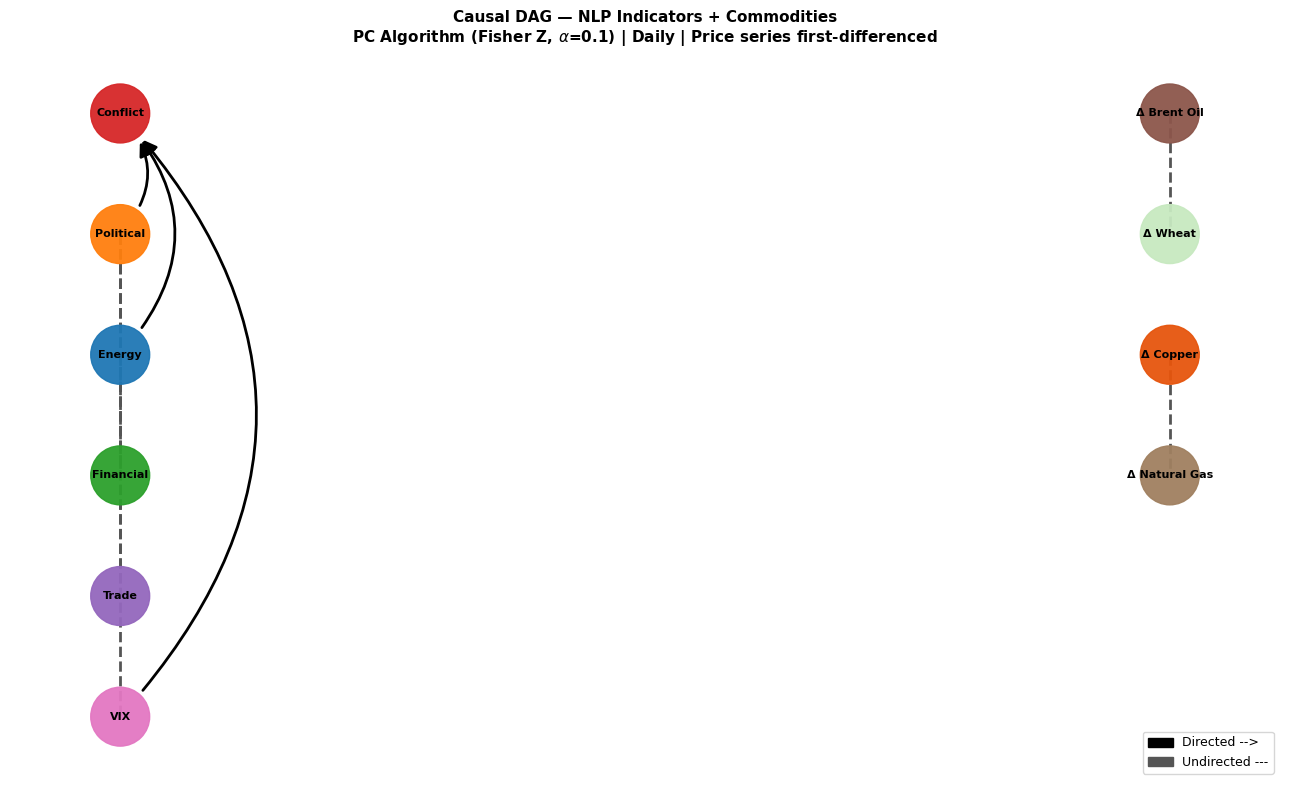}
\caption{Causal DAG 2 --- NLP indicators + commodity markets
($\Delta$Brent, $\Delta$Wheat, $\Delta$Copper,
$\Delta$Natural Gas) and the VIX.
PC algorithm, Fisher Z, $\alpha = 0.10$.
$N = 267$ daily observations.}
\label{fig:dag_commodities}
\end{figure}

\begin{figure}[!htbp]
\centering
\includegraphics[width=\linewidth,height=0.45\textheight,
    keepaspectratio]{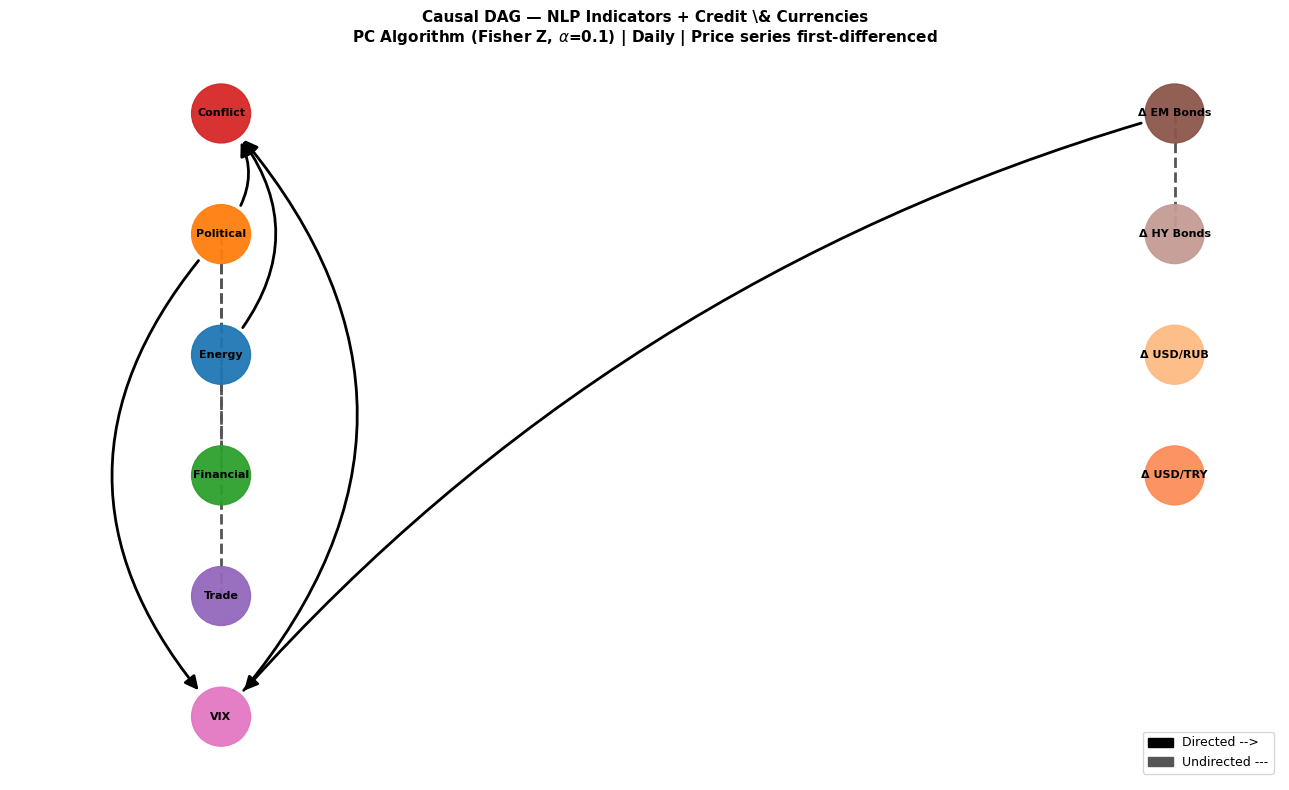}
\caption{Causal DAG 3 --- NLP indicators + credit and
currency markets ($\Delta$EM Bonds, $\Delta$HY Bonds,
$\Delta$USD/RUB, $\Delta$USD/TRY) and the VIX.
PC algorithm, Fisher Z, $\alpha = 0.10$.
$N = 267$ daily observations.}
\label{fig:dag_credit}
\end{figure}

\begin{figure}[!htbp]
\centering
\includegraphics[width=\linewidth,height=0.45\textheight,
    keepaspectratio]{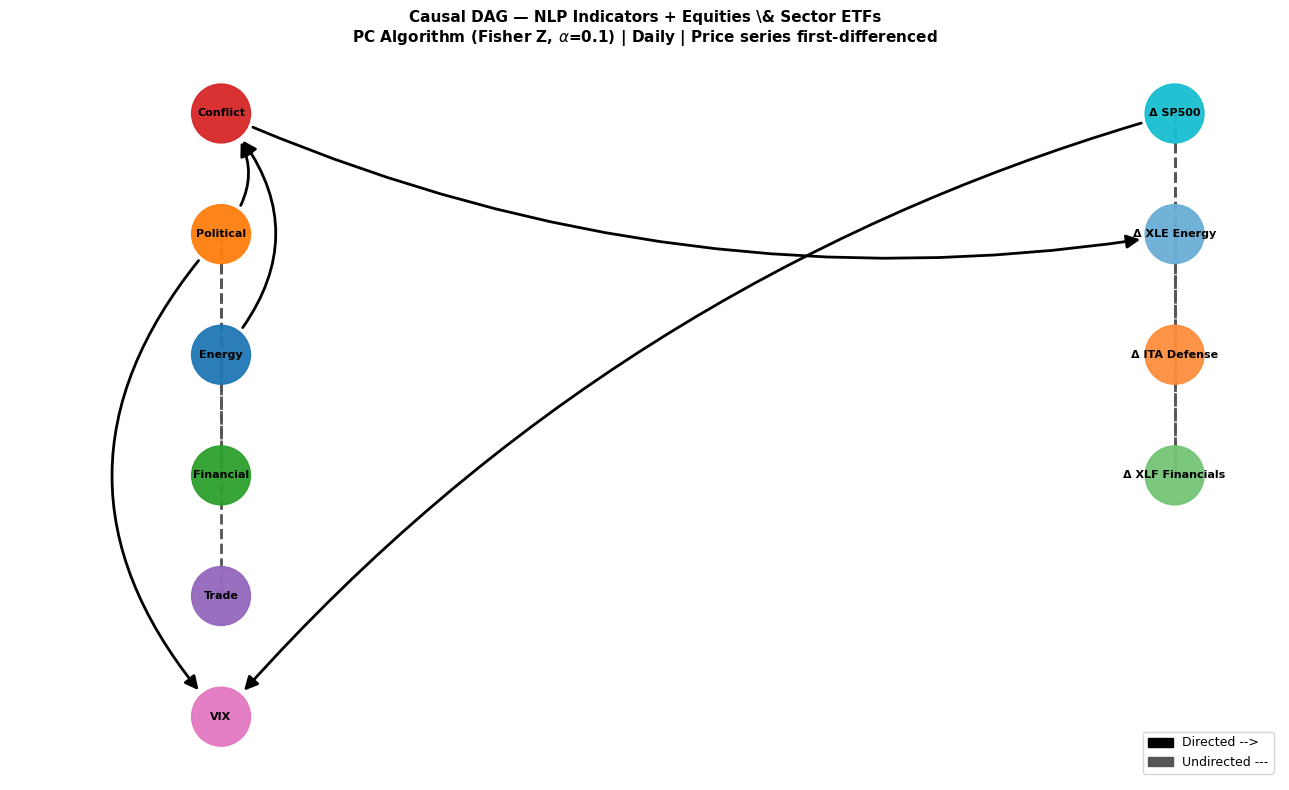}
\caption{Causal DAG 4 --- NLP indicators + equity sector ETFs
($\Delta$SP500, $\Delta$XLE energy, $\Delta$ITA defense,
$\Delta$XLF financials) and the VIX.
PC algorithm, Fisher Z, $\alpha = 0.10$.
Solid arrows: directed edges. Dashed lines: undirected edges.
$N = 267$ daily observations.
\textdagger\ The edge Conflict $\to$ $\Delta$XLE Energy is
plotted using the $\alpha = 0.05$ orientation, which we treat
as preferred on statistical and economic grounds; see
Section~\ref{sec:robustness}.}
\label{fig:dag_equities}
\end{figure}

Figures~\ref{fig:dag_core}--\ref{fig:dag_equities} report the
full CPDAGs for each of the four DAG specifications at the baseline
significance level $\alpha = 0.10$. In each figure, solid arrows
denote directed edges and dashed lines denote undirected edges whose
orientation the PC algorithm cannot determine from the data alone.

In DAG 1 (Figure~\ref{fig:dag_core}), the two globally robust
edges --- Political $\to$ Conflict, Energy $\to$ Conflict, and the near-robust edges
VIX $\to$ Conflict --- are all present, together with two
domain-specific directed edges. These are $\Delta$Gold $\to$ Financial
which implies that daily gold price movements causally precede financial
stress media coverage, consistent with gold acting as a leading
safe-haven barometer whose price reflects deteriorating financial
conditions before they are reported in the news cycle
\citep{BakerBloomDavis2016}. $\Delta$SP500 $\to$ VIX confirms
the well-established leverage effect in which equity market
declines drive implied volatility \citep{Kyle1985}. Energy
$\to$ Financial is also identified, suggesting that energy
stress narratives independently contribute to financial stress
coverage beyond the conflict channel. Three undirected edges
are present: Political $---$ Energy, Political $---$ Trade,
and $\Delta$Gold $---$ $\Delta$SP500. The first two reflect
strong associations between political instability, energy
security, and trade disruption narratives whose causal direction
cannot be identified from daily observational data --- a finding
consistent with these dimensions of geopolitical risk being
mutually reinforcing rather than sequentially ordered. The
undirected $\Delta$Gold $---$ $\Delta$SP500 edge reflects
the well-known negative correlation between equity returns
and gold prices in risk-off environments, which is
contemporaneous at daily frequency and therefore not
orientable by the PC algorithm.

DAG 2 (Figure~\ref{fig:dag_commodities}) confirms the two
globally robust NLP edges --- Political $\to$ Conflict and
Energy $\to$ Conflict --- and adds two notable undirected
commodity pairs. $\Delta$Brent $---$ $\Delta$Wheat reflects
the well-documented food-energy price co-movement
\citep{ChowdhuryEtAl2025}, amplified during the sample period
by the Russia--Ukraine conflict's simultaneous disruption of
grain and energy exports. $\Delta$Copper $---$ $\Delta$Natural
Gas captures the co-movement between industrial metals and
energy prices driven by common macroeconomic demand factors.
Neither pair can be oriented at daily frequency, consistent
with contemporaneous price discovery across commodity markets.
Notably, all four commodity market variables are completely
disconnected from the NLP indicators --- no directed or
undirected edge connects the right-hand commodity cluster
to the left-hand geopolitical cluster. This null result
reflects the sparsity of the financial NLP indicator
(0.2\% average message share) and the absence of
commodity-specific keywords in the OSINT channel
composition, rather than a true absence of geopolitical
transmission to commodity markets.

DAG 3 (Figure~\ref{fig:dag_credit}) yields one domain-specific
directed edge beyond the globally robust set: $\Delta$EM Bonds
$\to$ VIX, which survives at all three significance levels
including $\alpha = 0.05$ (bootstrap probability 0.330),
making it the most statistically conservative finding in the
entire analysis. This result implies that emerging market
sovereign bond price declines causally precede increases in
implied equity volatility --- consistent with EM sovereign
stress acting as an early warning signal of broader financial
market fear, a transmission consistent with the 2022 Russia
sanctions episode in which EM bond selloffs preceded the VIX
spike by several days. The undirected $\Delta$EM Bonds $---$
$\Delta$HY Bonds edge reflects the high correlation between
these two credit instruments, whose co-movement the algorithm
detects but cannot orient. Notably, $\Delta$USD/RUB and
$\Delta$USD/TRY are completely disconnected from both the
NLP indicators and VIX in DAG 3 --- daily ruble and lira
movements carry no incremental causal information about
geopolitical media coverage beyond what is already captured
by the NLP indicators. This null result is itself
informative: currency markets in geopolitically exposed
countries do not appear to lead the OSINT news cycle at
daily frequency.

DAG 4 (Figure~\ref{fig:dag_equities}) yields an
orientation-sensitive result between $\Delta$XLE Energy and
Conflict media coverage. At $\alpha = 0.05$ --- the most
statistically conservative threshold --- the PC algorithm
orients the edge as Conflict $\to$ $\Delta$XLE Energy,
consistent with the well-documented transmission from
geopolitical escalation to energy sector equity returns
\citep{CaldaraIacoviello2022, VerduzcoZanetti2026}.
At the looser thresholds $\alpha = 0.10$ and $\alpha = 0.15$
the orientation reverses, a pattern we attribute to reduced
discriminatory power of the independence tests at larger
conditioning sets rather than a genuine causal reversal.
We therefore treat Conflict $\to$ $\Delta$XLE Energy as
the preferred orientation, consistent with both the economic
prior and the strictest identification criterion. Beyond
this edge, DAG 4 confirms the globally robust findings:
Political $\to$ Conflict and Energy $\to$ Conflict appear
at all three significance levels, and $\Delta$SP500 $\to$
VIX is stable at $\alpha = 0.10$ and $\alpha = 0.15$.
The four sector ETFs form a tightly connected undirected
cluster at all significance levels, reflecting a common
equity factor structure driven by a shared market beta
that the CPDAG cannot orient given their high pairwise
correlations --- an expected result consistent with the
Arbitrage Pricing Theory \citep{Ross1976}.

\subsection{SVAR Impulse Responses}
\label{sec:svar}

To quantify the dynamic transmission between geopolitical risk
indicators and financial market variables, we estimate a
Structural VAR of order $p = 1$ --- selected by AIC and BIC
unanimously --- on the DAG 1 core macro panel. The Cholesky
ordering follows the topological sort of the directed edges
identified in DAG 1: $\Delta$SP500 $\to$ $\Delta$Gold $\to$
Political $\to$ Energy $\to$ Trade $\to$ VIX $\to$ Conflict
$\to$ Financial $\to$ $\Delta$Brent Oil, placing the most
exogenous variables first.

\begin{figure}[!htbp]
\centering
\includegraphics[width=\linewidth,height=0.72\textheight,
    keepaspectratio]{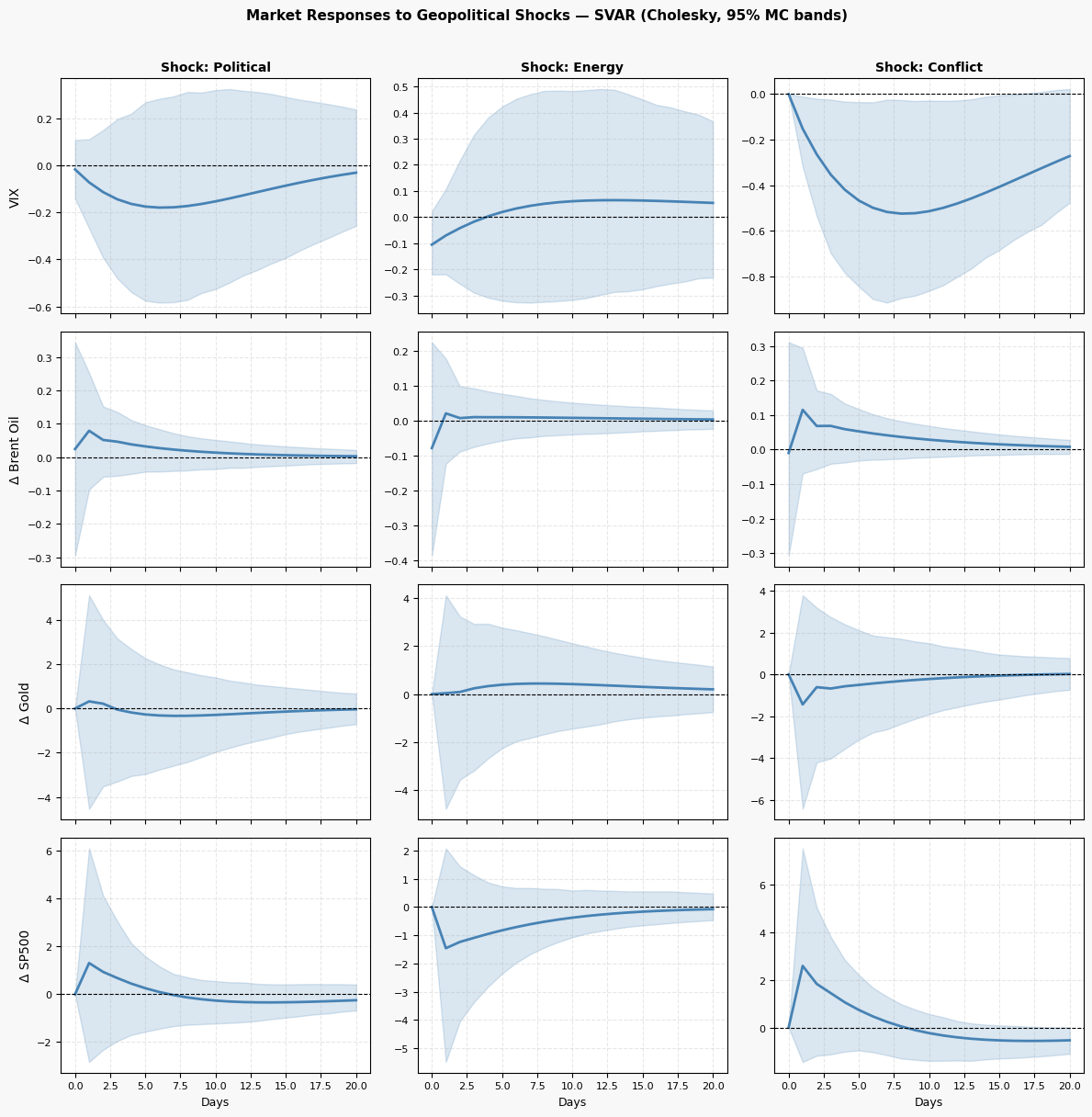}
\caption{Market responses to geopolitical shocks ---
SVAR (Cholesky, 95\% MC confidence intervals).
Columns: shock variables (Political, Energy, Conflict).
Rows: market responses (VIX, $\Delta$Brent Oil,
$\Delta$Gold, $\Delta$SP500).
Ordering: $\Delta$SP500 $\to$ $\Delta$Gold $\to$
Political $\to$ Energy $\to$ Trade $\to$ VIX $\to$
Conflict $\to$ Financial $\to$ $\Delta$Brent Oil.
Lag order $p = 1$ selected by AIC. $N = 267$ daily
observations.}
\label{fig:irf}
\end{figure}

Figure~\ref{fig:irf} reports impulse response functions for
the responses of market variables (VIX, $\Delta$Brent Oil,
$\Delta$Gold, $\Delta$SP500) to one-standard-deviation shocks
in the geopolitical NLP indicators (Political, Energy,
Conflict), with 95\% Monte Carlo confidence intervals based
on 1,000 replications. The predominant finding is that
confidence bands are wide and include zero for virtually all
geopolitical shock--market response pairs, indicating that
the daily transmission from NLP geopolitical signals to
financial market prices is statistically weak at this
frequency and horizon. This result is consistent with the
DAG analysis, where no directed edge from NLP indicators
to market price series survived the robustness criteria ---
geopolitical signals causally precede other NLP dimensions
(e.g., Conflict) but not market prices directly.

The one exception is the Conflict $\to$ VIX response, which
is negative and partially excludes zero at short horizons
(days 1--3). This suggests that a shock to conflict media
coverage is associated with a modest decrease in implied
volatility --- a counterintuitive result that may reflect
mean reversion dynamics in VIX following conflict escalation
episodes, or the Cholesky ordering placing Conflict upstream
of VIX which restricts the contemporaneous response.

\begin{figure}[!htbp]
\centering
\includegraphics[width=\linewidth,height=0.35\textheight,
    keepaspectratio]{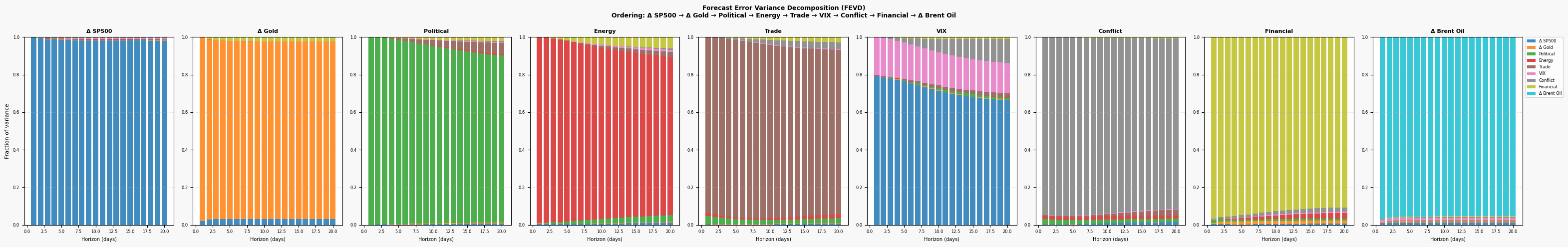}
\caption{Forecast error variance decomposition (FEVD).
Ordering: $\Delta$SP500 $\to$ $\Delta$Gold $\to$
Political $\to$ Energy $\to$ Trade $\to$ VIX $\to$
Conflict $\to$ Financial $\to$ $\Delta$Brent Oil.
Horizon: 20 trading days. $N = 267$ daily observations.}
\label{fig:fevd}
\end{figure}

Figure~\ref{fig:fevd} reports the forecast error variance
decompositions.

The FEVD confirms the limited explanatory power of geopolitical
NLP indicators for market price dynamics at daily frequency.
Own-variance dominates forecast error variance for all nine
variables throughout the 20-day horizon, with the geopolitical
NLP indicators explaining less than 5\% of the forecast
variance of VIX, $\Delta$Brent Oil, $\Delta$Gold, and
$\Delta$SP500 at all horizons. The Trade indicator exhibits
a modest contribution to VIX variance decomposition beyond
day 10, consistent with the directed Trade $\to$ VIX edge
identified in the DAG analysis. These results reinforce the
conclusion that geopolitical news signals operate primarily
within the media narrative system --- shaping how conflict,
political instability, and energy stress are covered ---
rather than generating statistically detectable short-run
price movements in financial markets at daily frequency.
This motivates the extension to weekly or monthly aggregation
as a direction for future work, where lower-frequency
transmission channels may be more precisely identified.

\subsection{Prediction Market Extension}
\label{sec:polymarket}

We extend the \CA{} framework using a deployed prediction-market
data layer based on Polymarket, the world's largest decentralised
prediction market. For a curated set of liquid event contracts
mapped to our five geopolitical risk categories --- covering
conflict escalation, energy supply disruption, political regime
change, financial sanctions, and trade restriction events --- we
capture daily point-in-time implied probabilities, 24-hour trading
volume, and period-over-period price changes, applying the quality
filters of \citet{WolfersZitzewitz2004}: minimum 24-hour trading
volume of \$250,000, bid-ask spread below 5 cents on a \$1.00
contract, and at least 14 days to contract resolution.

This prediction-market layer provides a third information signal
distinct from both the OSINT NLP indicators and asset prices: a
forward-looking, market-implied probability series reflecting the
aggregated beliefs of financially incentivised traders. Placing
this series in the DAG allows us to test directly whether
prediction market probabilities lead the OSINT news cycle ---
sharpening the paper's central finding on information ordering
--- and to locate prediction markets in the causal ordering
relative to both media coverage and asset prices.

The episode documented by \citet{PolymarketIran2026} is
illustrative: six wallets earned approximately \$1.2 million by
correctly betting on prediction market contracts tied to Iranian
military developments before the strikes appeared in news coverage,
suggesting that prediction markets may incorporate non-public
information faster than both OSINT channels and traditional media.
If this ordering is systematic, a three-layer causal structure ---
NLP indicators $\to$ prediction market probabilities $\to$ asset
prices --- would represent the full information chain from
geopolitical events to financial markets. We test this hypothesis
and report results in ongoing work.

% ============================================================
\section{Robustness}
\label{sec:robustness}
% ============================================================

We assess the stability of our findings across three dimensions.
First, we re-estimate all four DAGs at $\alpha \in \{0.05, 0.10,
0.15\}$. The two globally robust edges --- Political $\to$ Conflict
and Energy $\to$ Conflict --- are present in all four DAG
specifications at $\alpha = 0.10$ and $\alpha = 0.15$, but drop
out at $\alpha = 0.05$, reflecting the limited power of
independence tests at the effective sample size of approximately
38 bootstrap blocks. Second, block-bootstrap resampling (500
resamples, block size 7 days) yields probabilities of 0.330 and
0.252 for these two edges respectively, confirming moderate
but not strong resampling stability. The domain-specific edge $\Delta$EM Bonds
$\to$ VIX is the most statistically conservative finding,
surviving at $\alpha = 0.05$ with a bootstrap probability of
0.330. Third, the multi-DAG cross-specification criterion ---
requiring each edge to appear in every eligible DAG --- confirms
that the globally robust edges are not artefacts of a particular
asset class variable set. Complete edge lists at each significance
level are reported in Appendix~\ref{app:alpha_sweep}.

A comparison of the \CA{} Conflict indicator with the
\citet{CaldaraIacoviello2022} monthly \GPR{} index is left for
future work, as the \CA{} sample (April 2025--April 2026) only
partially overlaps with the period for which the GPR index
provides reliable monthly variation. We note that the two
indicators are conceptually distinct --- the GPR index aggregates
newspaper coverage of geopolitical risk broadly defined, while
\CA{} focuses specifically on OSINT Telegram channels with a
conflict-monitoring orientation --- and a meaningful comparison
would require a longer overlapping sample.

% ============================================================
\section{Conclusion}
\label{sec:conclusion}
% ============================================================

This paper introduces \CA{}, an open-source framework for real-time
geopolitical risk measurement and causal analysis of financial market
dynamics. Our Telegram-based \GPR{} index provides a high-frequency
alternative to newspaper-based measures, decomposing geopolitical
signals into five thematically distinct indicators across six OSINT
channels. The application of the PC algorithm across four DAG
specifications and three significance levels yields a consistent and
economically interpretable causal architecture.

Two findings emerge as globally robust. Political instability and
energy media coverage independently and causally precede conflict
coverage in all four DAG specifications, establishing conflict as
the primary causal sink of geopolitical narrative escalation in
real-time conflict-oriented OSINT channels. At the strictest significance level
($\alpha = 0.05$), conflict coverage causally precedes $\Delta$XLE
energy sector returns, consistent with geopolitical escalation
transmitting to energy equity markets \citep{CaldaraIacoviello2022}.
The SVAR analysis indicates that transmission from geopolitical
NLP signals to financial market prices is statistically weak at
daily frequency, suggesting that geopolitical news signals operate
primarily within the media narrative system rather than generating
detectable short-run price movements.

These findings have practical implications for sovereign risk
monitoring and early-warning systems: systematic monitoring of
political and energy narratives in OSINT channels may provide
leading signals of conflict escalation before it appears in
traditional geopolitical risk indices. Future work will extend
the analysis to lower frequencies where transmission to financial
markets may be more precisely identified, incorporate prediction
market probabilities as a third information layer between media
narratives and asset prices, and apply rolling-window estimation
to assess causal structure stability across geopolitical regimes.

% ============================================================
\clearpage
\bibliographystyle{chicago}
\bibliography{references}

% ============================================================
\clearpage
\subsection*{Appendix A: GPR Keyword Dictionary}
\addcontentsline{toc}{subsection}{Appendix A: GPR Keyword Dictionary}
\label{app:keywords}

\begin{table}[!htbp]
\centering
\caption{Keyword categories and selected examples}
\label{tab:keywords}
\begin{tabular}{L{3.5cm} L{9.5cm}}
\toprule
\textbf{Category} & \textbf{Keywords (selected)} \\
\midrule
Conflict \& security &
  war, invasion, airstrike, missile, ceasefire, drone,
  troops, casualties, terrorism, assassination, cyberattack \\
Political instability &
  coup, regime change, civil unrest, election fraud, protest,
  crackdown, martial law, authoritarian, political crisis \\
Energy security &
  natural gas, lng, pipeline, energy crisis, supply disruption,
  strait of hormuz, tanker, refinery, energy embargo \\
Financial stress &
  default, debt crisis, sovereign debt, bank run, central bank,
  inflation, devaluation, currency crisis, market crash \\
Trade disruption &
  tariff, trade war, export controls, supply chain, shortage,
  semiconductor, rare earth, grain, shipping, logistics \\
\bottomrule
\end{tabular}
\end{table}

\subsection*{Appendix B: Alpha Sweep Supplementary Results}
\addcontentsline{toc}{subsection}{Appendix B: Alpha Sweep Supplementary Results}
\label{app:alpha_sweep}

Tables~\ref{tab:alpha_dag1}--\ref{tab:alpha_dag4} report the
complete directed edge lists for each DAG specification at
$\alpha \in \{0.05, 0.10, 0.15\}$.

\begin{table}[!htbp]
\centering
\caption{DAG 1 (Core Macro) --- directed edges by significance level}
\label{tab:alpha_dag1}
\begin{threeparttable}
\begin{tabular}{llccc}
\toprule
\textbf{Source} & \textbf{Target} & $\alpha=0.05$ & $\alpha=0.10$ & $\alpha=0.15$ \\
\midrule
Political          & Conflict    & \checkmark & \checkmark & \checkmark \\
Energy             & Conflict    & \checkmark & \checkmark & \checkmark \\
VIX                & Conflict    &            & \checkmark & \checkmark \\
Political          & VIX         &            & \checkmark & \checkmark \\
Energy             & Financial   &            & \checkmark & \checkmark \\
$\Delta$ Gold      & Financial   &            & \checkmark & \checkmark \\
$\Delta$ SP500     & VIX         &            & \checkmark & \checkmark \\
Conflict           & $\Delta$ XLE Energy & \checkmark &   &            \\
Political          & Energy      & \checkmark &            &            \\
Financial          & Energy      & \checkmark &            &            \\
$\Delta$ XLF Fin.  & $\Delta$ XLE Energy & \checkmark &  &            \\
\bottomrule
\end{tabular}
\begin{tablenotes}
\item \textit{Note:} \checkmark\ indicates the directed edge is
present at that significance level. Undirected edges omitted
for brevity.
\end{tablenotes}
\end{threeparttable}
\end{table}

\begin{table}[!htbp]
\centering
\caption{DAG 2 (Commodities) --- directed edges by significance level}
\label{tab:alpha_dag2}
\begin{threeparttable}
\begin{tabular}{llccc}
\toprule
\textbf{Source} & \textbf{Target} & $\alpha=0.05$ & $\alpha=0.10$ & $\alpha=0.15$ \\
\midrule
Political      & Conflict  & \checkmark & \checkmark & \checkmark \\
Energy         & Conflict  & \checkmark & \checkmark & \checkmark \\
VIX            & Conflict  &            & \checkmark & \checkmark \\
\bottomrule
\end{tabular}
\begin{tablenotes}
\item \textit{Note:} DAG 2 yields fewer directed edges than other
specifications, consistent with commodity price changes being
predominantly explained by own dynamics at daily frequency.
\end{tablenotes}
\end{threeparttable}
\end{table}

\begin{table}[!htbp]
\centering
\caption{DAG 3 (Credit \& FX) --- directed edges by significance level}
\label{tab:alpha_dag3}
\begin{threeparttable}
\begin{tabular}{llccc}
\toprule
\textbf{Source} & \textbf{Target} & $\alpha=0.05$ & $\alpha=0.10$ & $\alpha=0.15$ \\
\midrule
Political          & Conflict   & \checkmark & \checkmark & \checkmark \\
Energy             & Conflict   & \checkmark & \checkmark & \checkmark \\
VIX                & Conflict   &            & \checkmark & \checkmark \\
Political          & VIX        &            & \checkmark & \checkmark \\
$\Delta$ EM Bonds  & VIX        & \checkmark & \checkmark & \checkmark \\
\bottomrule
\end{tabular}
\begin{tablenotes}
\item \textit{Note:} $\Delta$EM Bonds $\to$ VIX is the only
domain-specific edge that survives at all three significance
levels, making it the most statistically conservative
domain-specific finding in the analysis.
\end{tablenotes}
\end{threeparttable}
\end{table}

\begin{table}[!htbp]
\centering
\caption{DAG 4 (Sector ETFs) --- directed edges by significance level}
\label{tab:alpha_dag4}
\begin{threeparttable}
\begin{tabular}{llccc}
\toprule
\textbf{Source} & \textbf{Target} & $\alpha=0.05$ & $\alpha=0.10$ & $\alpha=0.15$ \\
\midrule
Political              & Conflict  &            & \checkmark & \checkmark \\
Energy                 & Conflict  &            & \checkmark & \checkmark \\
Political              & VIX       &            & \checkmark & \checkmark \\
$\Delta$ SP500         & VIX       &            & \checkmark & \checkmark \\
Conflict               & $\Delta$ XLE Energy & \checkmark &  &   \\
$\Delta$ XLE Energy    & Conflict  &            & \checkmark & \checkmark \\
Political              & Energy    & \checkmark &            &            \\
Financial              & Energy    & \checkmark &            &            \\
$\Delta$ XLF Fin.      & $\Delta$ XLE Energy & \checkmark & & \\
\bottomrule
\end{tabular}
\begin{tablenotes}
\item \textit{Note:} The orientation reversal between Conflict
and $\Delta$XLE Energy across significance levels is discussed
in Section~\ref{sec:dag_results}. We treat the $\alpha = 0.05$
orientation (Conflict $\to$ $\Delta$XLE Energy) as preferred
on statistical and economic grounds.
\end{tablenotes}
\end{threeparttable}
\end{table}

\end{document}